\documentclass[fleqn,12pt]{article}

\usepackage{latexsym,ifthen,epsfig,color}
\usepackage{amsmath,amssymb,amsthm}
\usepackage{float}

\newcommand{\join}{\text{\textcircled{{\footnotesize 1}}}}
\newcommand{\cojoin}{\text{\textcircled{{\footnotesize 0}}}}

\newcommand{\NP}{\ensuremath{\mathbb{NP}}}

\newtheorem{clai}{Claim}[section]
\newtheorem{theo}{Theorem}
\newtheorem{lemma}{Lemma}

\newtheorem{coro}{Corollary}

\newtheorem{defi}{Definition}

\begin{document}

\author{
Andreas Brandst\"adt\\
\small Institut f\"ur Informatik, Universit\"at Rostock, D-18051 Rostock, Germany\\
\small \texttt{andreas.brandstaedt@uni-rostock.de}\\
\and
Raffaele Mosca\\
\small Dipartimento di Economia, Universit\'a degli Studi ``G. D'Annunzio'', 
Pescara 65121, Italy\\
\small \texttt{r.mosca@unich.it}
}

\title{Finding Efficient Domination for $S_{1,1,5}$-Free Bipartite Graphs in Polynomial Time}

\maketitle

\begin{abstract}
A vertex set $D$ in a finite undirected graph $G$ is an {\em efficient dominating set} (e.d.s.\ for short) of $G$ if every vertex of $G$ is dominated by exactly one vertex of $D$. The \emph{Efficient Domination} (ED) problem, which asks for the existence of an e.d.s.\ in $G$, is \NP-complete for various $H$-free bipartite graphs, e.g., Lu and Tang showed that ED is \NP-complete for chordal bipartite graphs and for planar bipartite graphs; actually, ED is \NP-complete even for planar bipartite graphs with vertex degree at most 3 and girth at least $g$ for every fixed $g$. Thus, ED is \NP-complete for $K_{1,4}$-free bipartite graphs and for $C_4$-free bipartite graphs. 

In this paper, we show that ED can be solved in polynomial time for $S_{1,1,5}$-free bipartite graphs. 
\end{abstract}


\section{Introduction}\label{sec:intro}

Let $G=(V,E)$ be a finite undirected graph. A vertex $v$ {\em dominates} itself and its neighbors. A vertex subset $D \subseteq V$ is an {\em efficient dominating set} ({\em e.d.s.}\ for short) of $G$ if every vertex of $G$ is dominated by exactly one vertex in $D$; for any e.d.s.\ $D$ of $G$, $|D \cap N[v]| = 1$ for every $v \in V$ (where $N[v]$ denotes the closed neighborhood of $v$).
Note that not every graph has an e.d.s.; the {\sc Efficient Dominating Set} (ED for short) problem asks for the existence of an e.d.s.\ in a given graph~$G$.

\medskip

In \cite{BanBarSla1988,BanBarHosSla1996}, it was shown that the ED problem is \NP-complete, and it is \NP-complete for $P_k$-free graphs, $k \ge 7$. 
However, in \cite{BraMos2016}, we have shown that ED is solvable in polynomial time for $P_6$-free graphs which leads to a dichotomy of ED for $H$-free graphs. 

\medskip

Lu and Tang \cite{LuTan2002} showed that ED is \NP-complete for chordal bipartite graphs (i.e., hole-free bipartite graphs). Thus, for every $k \ge 3$, ED is \NP-complete for $C_{2k}$-free bipartite graphs. 
Moreover, ED is \NP-complete for planar bipartite graphs \cite{LuTan2002} and even for planar bipartite graphs of maximum degree 3 \cite{BraMilNev2013} and girth at least $g$ for every fixed $g$ \cite{Nevri2014}. Thus, ED is \NP-complete for $K_{1,4}$-free bipartite graphs and for $C_4$-free bipartite graphs.

\medskip

It is well known that for every graph class with bounded clique-width, ED can be solved in polynomial time \cite{CouMakRot2000}; for instance, the clique-width of  claw-free bipartite, i.e., $K_{1,3}$-free bipartite graphs is bounded. Dabrowski and Paulusma  \cite{DabPau2016} published a dichotomy for clique-width of $H$-free bipartite graphs. For instance, the clique-width of $S_{1,1,3}$-free bipartite graphs is bounded (which includes $K_{1,3}$-free bipartite graphs). 
However, the clique-width of $S_{1,1,5}$-free bipartite graphs is unbounded. 

\medskip

In \cite{BraMos2019}, we solved ED in polynomial time for $P_7$-free bipartite graphs, for $\ell P_4$-free bipartite graphs for fixed $\ell$, for $S_{2,2,4}$-free bipartite graphs as well as for $P_9$-free bipartite graphs with degree at most 3, but we had some open problems: 
What is the complexity of ED for 
\begin{itemize}
\item[$-$] $P_k$-free bipartite graphs, $k \ge 8$, 
\item[$-$] $S_{1,3,3}$-free bipartite graphs, 
\item[$-$] $S_{1,1,5}$-free bipartite graphs,
\item[$-$] $S_{2,2,k}$-free bipartite graphs for $k \ge 5$,
\item[$-$] chordal bipartite graphs with vertex degree at most 3? 
\end{itemize}

In \cite{BraMos2020}, we showed already that for $S_{1,3,3}$-free bipartite graphs, ED can be solved in polynomial time.
In this manuscript, we will show:

\begin{theo}\label{EDS115frbipgr}
For $S_{1,1,5}$-free bipartite graphs, ED can be solved in polynomial time.
\end{theo}

\section{Preliminaries}

Let $G=(X,Y,E)$, $V(G)=X \cup Y$, be an $S_{1,1,5}$-free bipartite graph with black $X$ and white $Y$.
For a vertex set $U \subseteq V(G)$, a vertex $v \notin U$ has a {\em join} to $U$, say $v \join U$, if $v$ contacts every vertex in $U$.
Moreover, $v$ has a {\em co-join} to $U$, say $v \cojoin U$, if $v$ does not contact any vertex in $U$. 
Correspondingly, for vertex sets $U,W \subseteq V(G)$ with $U \cap W = \emptyset$,
$U \join W$ denotes $u \join W$ for all $u \in U$, and $U \cojoin W$ denotes $u \cojoin W$ for all $u \in U$. 
A vertex $u \notin W$ {\em contacts $W$} if $u$ has a neighbor in $W$. For vertex sets $U,W$ with $U \cap W = \emptyset$, $U$ {\em contacts $W$} if there is a vertex in $U$ contacting $W$.

\medskip

Let $dist_G(u,v)$ denote the minimum distance between $u$ and $v$ in $G$. 

\medskip

A vertex $u \in V$ is {\em forced} if $u \in D$ for every e.d.s.\ $D$ of $G$; $u$ is {\em excluded} if $u \notin D$ for every e.d.s.\ $D$ of $G$. For example, if $x_1,x_2 \in X$ are leaves in $G$ and $y$ is the neighbor of $x_1,x_2$ then $x_1,x_2$ are excluded and $y$ is forced.
By a forced vertex, $G$ can be reduced to $G'$ as follows:

\begin{clai}\label{forcedreduction}
If $u$ is forced then $G$ has an e.d.s.\ $D$ with $u \in D$ if and only if the reduced graph $G'=G \setminus N[u]$ has an e.d.s.\ $D'=D \setminus \{u\}$ such that all vertices in $N^2(u)$ are excluded in $G'$.
\end{clai}

Analogously, if we assume that $v \in D$ for a vertex $v \in V=X \cup Y$ then $u \in V$ is {\em $v$-forced} if $u \in D$ for every e.d.s.\ $D$ of $G$ with $v \in D$,
and $u$ is {\em $v$-excluded} if $u \notin D$ for every e.d.s.\ $D$ of $G$ with $v \in D$. For checking whether $G$ has an e.d.s.\ $D$ with $v \in D$, we can clearly reduce $G$ by forced vertices as well as by $v$-forced vertices when we assume that $v \in D$:

\begin{clai}\label{vforcedreduction}
If we assume that $v \in D$ and $u$ is $v$-forced then $G$ has an e.d.s.\ $D$ with $v \in D$ if and only if the reduced graph $G'=G \setminus N[u]$ has an e.d.s.\ $D'=D \setminus \{u\}$ with $v \in D'$ such that all vertices in $N^2(u)$ are $v$-excluded in $G'$.
\end{clai}

Similarly, for $k \ge 2$, $u \in V$ is {\em $(v_1,\ldots,v_k)$-forced} if $u \in D$ for every e.d.s.\ $D$ of $G$ with $v_1,\ldots,v_k \in D$, and correspondingly, $u \in V$ is {\em $(v_1,\ldots,v_k)$-excluded} if $u \notin D$ for such e.d.s.\ $D$, and $G$ can be reduced by the same principle.

\medskip

Clearly, for every component of $G$, the e.d.s.\ problem is independently solvable. Thus, we can assume that $G$ is connected.
Since $G$ is $S_{1,1,5}$-free bipartite, $G$ is not really $P_k$-free for every $k \ge 1$. 

\medskip

When no $D$-vertex is midpoint of a $P_4$ in $G$ then every midpoint of $P_4$ is excluded, and every endpoint of $P_4$ in $G$ is forced or there is no e.d.s.\ in $G$ if two endpoints of $P_4$ have a common neighbor. In this case, the e.d.s.\ problem can be easily done in polynomial time.  
Thus, from now on, assume that there exists a $D$-vertex which is midpoint of a $P_4$. 

\medskip

By the e.d.s.\ property, for every $v,w \in D$, $dist_G(v,w) \ge 3$. Moreover, for $v,w \in D \cap X$ or $v,w \in D \cap Y$, $dist_G(v,w) \ge 4$.   
Assume that there exist $D$-vertices $x,y \in D$ with $dist_G(x,y)=3$, and $x,y \in D_{basis}$. 
Let $N_0:=D_{basis}$ with $x,y \in N_0$ and moreover, every $(x,y)$-forced vertex is in $N_0$. Let $N_i$, $i \ge 1$, be the distance levels of $D_{basis}$, 
i.e., $N_i := \{v \in X \cup Y:$ min $\{dist_G(v,d): d \in D_{basis}\} = i\}$, $i \ge 1$.  

\medskip

By the e.d.s.\ property, we have:
\begin{equation}\label{(1)}
D \cap (N_1 \cup N_2)=\emptyset.
\end{equation}

If there is a vertex $u \in N_2$ with $N(u) \cap N_3= \emptyset$ then by the e.d.s.\ property, $G$ has no e.d.s.\ $D$ with such $D_{basis}$. Thus, we assume:
\begin{equation}\label{(2)}
\mbox{ Every vertex } u \in N_2 \mbox{ has a neighbor in } N_3.
\end{equation}

If there is a vertex $u \in N_2$ with $|N(u) \cap N_3|=1$, say $N(u) \cap N_3=\{w\}$ then $w$ is $D_{basis}$-forced. Moreover, if there are 
$u,u' \in N_2$ with $N(u) \cap N_3=\{w\}$, $N(u') \cap N_3=\{w'\}$, $w \neq w'$, and $dist_G(w,w')=1$ or $dist_G(w,w')=2$ then by the e.d.s.\ property, $G$ has no e.d.s.\ $D$ with such $D_{basis}$.  
Thus assume that for $N(u) \cap N_3=\{w\}$, $N(u') \cap N_3=\{w'\}$, $dist_G(w,w') \ge 3$.  

\medskip

Then for every $u \in N_2$ with $N(u) \cap N_3=\{w\}$, one can update $D_{basis}$, i.e., $D_{basis}:= D_{basis} \cup \{w\}$ and redefine the distance levels $N_i$, $i \ge 1$, with respect to $D_{basis}$. 
Thus, we assume:
\begin{equation}\label{(3)}
\mbox{ Every vertex } u \in N_2 \mbox{ has at least two neighbors in } N_3.
\end{equation}

If there is a vertex $w \in N_3$ with $N(w) \cap (N_3 \cup N_4)=\emptyset$ then $w$ is $D_{basis}$-forced. 
Moreover, if there are two such $w,w'$ with common neighbor $u \in N_2$, $uw \in E$, $uw' \in E$, and $N(w) \cap (N_3 \cup N_4)=\emptyset$ as well as 
$N(w') \cap (N_3 \cup N_4)=\emptyset$, then by the e.d.s.\ property, $G$ has no e.d.s.\ $D$ with such $D_{basis}$.  
Thus assume that for such $D_{basis}$-forced vertices $w,w' \in N_3$, there is no common neighbor in $N_2$. 

\medskip

Then for every $w \in N_3$ with $N(w) \cap (N_3 \cup N_4)=\emptyset$, one can update $D_{basis}$, i.e., $D_{basis}:= D_{basis} \cup \{w\}$ and redefine the distance levels 
$N_i$, $i \ge 1$, with respect to $D_{basis}$ as above. 
Thus, we assume:
\begin{equation}\label{(4)}
\mbox{ Every vertex } w \in N_3 \mbox{ has a neighbor in } N_3 \cup N_4.
\end{equation}

In particular, we have:

\begin{clai}\label{uinN2endpointP6winN3forced}
If $u \in N_2$ is the endpoint of a $P_6$ whose remaining vertices are in $N_0 \cup N_1$, and $uw \in E$ with $w \in N_3$ then $w$ is $D_{basis}$-forced.
\end{clai}

\noindent
{\bf Proof.} 
Assume that $u$ is the endpoint of a $P_6$ $(u,v_1,v_2,v_3,v_4,v_5)$ with $v_1 \in N_1$ and $v_2,v_3,v_4,v_5 \in N_0 \cup N_1$. If $|N(u) \cap N_3| \ge 2$, say 
$w,w' \in N(u) \cap N_3$, then $(u,w,w',v_1,v_2,v_3,v_4,v_5)$ (with midpoint $u$) induce an $S_{1,1,5}$, which is a contradiction. Thus, $|N(u) \cap N_3| = 1$, say $N(u) \cap N_3 = \{w\}$, and then $w$ is $D_{basis}$-forced.    
\qed

\medskip

Then one can update $D_{basis}$, i.e., $D_{basis}:=D_{basis} \cup \{w\}$. Thus, from now on, assume that every $u \in N_2$ is the endpoint of a $P_k$, $k \le 5$, but not the endpoint of a $P_6$, whose remaining vertices are in $N_0 \cup N_1$.
 
\begin{clai}\label{uinN2endpointP5winN3}
If $u \in N_2$ is the endpoint of a $P_5$ whose remaining vertices are in $N_0 \cup N_1$, and $uw \in E$ with $w \in N_3$, 
then $|N(w) \cap (N_3 \cup N_4)|=1$.  
\end{clai}

\noindent
{\bf Proof.}
Assume that $u$ is the endpoint of a $P_5$ $(u,v_1,v_2,v_3,v_4)$ with $v_1 \in N_1$ and $v_2,v_3,v_4 \in N_0 \cup N_1$. 
Clearly, by (\ref{(2)}), $u$ must have a $D$-neighbor $w \in N_3$. Recall that by (\ref{(4)}), $|N(w) \cap (N_3 \cup N_4)| \ge 1$.
If $|N(w) \cap (N_3 \cup N_4)| \ge 2$, say $wr_1 \in E$, $wr_2 \in E$, with $r_1, r_2 \in N_3 \cup N_4$ then
$(w,r_1,r_2,u,v_1,v_2,v_3,v_4)$ (with midpoint $w$) induce an $S_{1,1,5}$, which is a contradiction.
Thus, $|N(w) \cap (N_3 \cup N_4)|=1$.  
\qed

\section{When $(v_1,\ldots,v_6)$ is a $P_6$ or $C_6$ in $G$ with $v_2,v_5 \in D$}\label{2DP6C6}

In this section, we assume that there is a $P_6$ or $C_6$ $(v_1,v_2,v_3,v_4,v_5,v_6)$ in $G$ with $v_iv_{i+1} \in E$, $1 \le i \le 5$, and two $D$-vertices 
$v_2,v_5 \in D$, i.e., either 
\begin{itemize}
\item[$(i)$] $P=(v_1,\ldots,v_6)$ induce a $P_6$ in $G$, or
\item[$(ii)$] $C=(v_1,\ldots,v_6)$ induce a $C_6$ in $G$ with $v_6v_1 \in E$.
\end{itemize}

We call $P$ a $2D-P_6$ and $C$ a $2D-C_6$. Clearly, $G$ contains polynomially many induced $P_6$ and $C_6$. 
Without loss of generality, assume that $v_1,v_3,v_5 \in X$ and $v_2,v_4,v_6 \in Y$. 
By an iterative application of the above operations, we have:

\begin{clai}\label{Dbasis}
There is a set $D_{basis}$, with $\{v_2,v_5\} \subseteq D_{basis}$, such that:
\begin{itemize}
\item[$(i)$] Every vertex in $D_{basis} \setminus \{v_2,v_5\}$ is $(v_2,v_5)$-forced.
\item[$(ii)$] $G[D_{basis} \cup N(D_{basis})]$ is connected.
\item[$(iii)$] Assumptions $(\ref{(1)}),(\ref{(2)}),(\ref{(3)}),(\ref{(4)})$ hold for the distance levels $N_i$ with respect to $D_{basis}$, namely, 
$N_i := \{v \in X \cup Y: \min \{dist(v,d): d \in D_{basis}\} = i\}$, $i \ge 1$, and $N_0:=D_{basis}$.
\end{itemize}
In particular, $D_{basis}$ can be computed in polynomial time. \qed
\end{clai}

In what follows let $N_0=D_{basis}$ be just the set of Claim \ref{Dbasis} [i.e. let us assume that such a set has been computed] and let $N_i$, $i \geq 1$, 
be the corresponding distance levels of $D_{basis}$.

\subsection{General remarks}\label{genrem}

\begin{clai}\label{uinN2contactsforcedneighbor}
If $u \in N_2 \cap X$ contacts $y \in N(x)$ for a $(v_2,v_5)$-forced vertex $x \in N_0 \cap X$ then $u$ contacts $N(v_5)$. 
Analogously, if $u \in N_2 \cap Y$ contacts $x \in N(y)$ for a $(v_2,v_5)$-forced vertex $y \in N_0 \cap Y$ then $u$ contacts $N(v_2)$.  
\end{clai}

\noindent
{\bf Proof.} 
Without loss of generality, assume that $u \in N_2 \cap X$ contacts $y \in N(x)$ for a $(v_2,v_5)$-forced vertex $x \in N_0 \cap X$. Since it is a component with 
$N[v_2]$ and $N[v_5]$ and $y$ is white, we assume without loss of generality that $y$ contacts $N(v_2)$. 
Clearly, $u$ is black and $u$ does not contact $N(v_2)$. Moreover, $u$ must have a $D$-neighbor in $D \cap N_3$, by (\ref{(3)}), $u$ has at least two neighbors 
$r,r' \in N_3$, say $r \in D \cap N_3$, and by (\ref{(4)}), $r$ has a neighbor $s \in N_3 \cup N_4$. 

\medskip

Suppose to the contrary that $u$ does not contact $N(v_5)$. 

\medskip

First assume that $yv_1 \notin E$ and $yv_3 \notin E$, say $yv \in E$ with $v \in N(v_2)$, $v \neq v_1,v_3$. 
Then $(v_2,v_1,v_3,v,y,u,r,s)$ (with midpoint $v_2$) induce an $S_{1,1,5}$, which is a contradiction. 
Thus, $yv_1 \in E$ or $yv_3 \in E$. 

\medskip

\noindent
{\bf Case 1.} $P=(v_1,\ldots,v_6)$ is a $2D-P_6$. 

\medskip

Since $(u,r,r',y,v_3,v_4,v_5,v_6)$ (with midpoint $u$) does not induce an $S_{1,1,5}$, we have $yv_3 \notin E$.
Thus, $yv_1 \in E$. But then $(u,r,r',y,v_1,v_2,v_3,v_4)$ (with midpoint $u$) induce an $S_{1,1,5}$, which is a contradiction.
Thus, in this case, $u$ contacts $N(v_5)$.

\medskip

\noindent
{\bf Case 2.} $C=(v_1,\ldots,v_6)$ is a $2D-C_6$ with $v_6v_1 \in E$.

\medskip

Recall that $yv_1 \in E$ or $yv_3 \in E$. Since $(u,r,r',y,v_1,v_6,v_5,v_4)$ (with midpoint $u$) does not induce an $S_{1,1,5}$, we have
$yv_1 \notin E$ and thus, $yv_3 \in E$. But then $(u,r,r',y,v_3,v_4,v_5,v_6)$ (with midpoint $u$) induce an $S_{1,1,5}$, which is a contradiction.
Thus, again, in this case, $u$ contacts $N(v_5)$.

\medskip

Analogously, if white $u \in N_2$ contacts $x \in N(y)$ for a $(v_2,v_5)$-forced vertex $y \in Y$ then $u$ contacts $N(v_2)$.  
Thus, Claim \ref{uinN2contactsforcedneighbor} is shown. 
\qed

\begin{clai}\label{2DP6uinN2uv3oruv4}
Let $P=(v_1,\ldots,v_6)$ be a $2D$-$P_6$ with $v_2,v_5 \in D_{basis}$. 
\begin{itemize}
\item[$(i)$] If $u \in N_2 \cap Y$ and $uv_1 \in E$ then $uv_3 \in E$. Moreover, if $uv_3 \notin E$ then $uv_1 \notin E$ and for every $v \in N(v_2) \setminus \{v_1,v_3\}$ with $uv \in E$, $vv_4 \in E$.
\item[$(ii)$] If $u \in N_2 \cap X$ and $uv_6 \in E$ then $uv_4 \in E$. Moreover, if $uv_4 \notin E$ then $uv_6 \notin E$ and for every $v \in N(v_5) \setminus \{v_4,v_6\}$ with $uv \in E$, $vv_3 \in E$.  
\end{itemize}
\end{clai}

\noindent
{\bf Proof.}
$(i)$: Recall that by (\ref{(3)}), $|N(u) \cap N_3| \ge 2$, say $w,w' \in N(u) \cap N_3$, and recall that by Claim \ref{uinN2contactsforcedneighbor}, 
$u \in N_2 \cap Y$ contacts $N(v_2)$.  

First assume that $uv_1 \in E$. Since $(u,w,w',v_1,v_2,v_3,v_4,v_5)$ (with midpoint $u$) does not induce an $S_{1,1,5}$, we have $uv_3 \in E$. 

Now assume that $uv_3 \notin E$. Then $uv_1 \notin E$, i.e., $uv \in E$ for $v \in N(v_2) \setminus \{v_1,v_3\}$. 
Since $(u,w,w',v,v_2,v_3,v_4,v_5)$ (with midpoint $u$) does not induce an $S_{1,1,5}$, we have $vv_4 \in E$. 
Thus, $(i)$ is shown, and analogously, $(ii)$ can be shown in the same way. 
Thus, Claim \ref{2DP6uinN2uv3oruv4} is shown. 
\qed

\begin{clai}\label{2DC6uinN2uv3oruv4}
Let $C=(v_1,\ldots,v_6)$ be a $2D$-$C_6$ with $v_2,v_5 \in D_{basis}$. 
\begin{itemize}
\item[$(i)$] If $u \in N_2 \cap Y$ then either $uv_1 \in E$ and $uv_3 \in E$ or $uv_1 \notin E$ and $uv_3 \notin E$, 
and for every $v \in N(v_2) \setminus \{v_1,v_3\}$ with $uv \in E$, we have $vv_4 \in E$ and $vv_6 \in E$.
\item[$(ii)$] If $u \in N_2 \cap X$ then either $uv_4 \in E$ and $uv_6 \in E$ or $uv_4 \notin E$ and $uv_6 \notin E$ 
and for every $v \in N(v_2) \setminus \{v_4,v_6\}$ with $uv \in E$, we have $vv_1 \in E$ and $vv_3 \in E$. 
\end{itemize}
\end{clai}

\noindent
{\bf Proof.}
$(i)$: Recall that by (\ref{(3)}), $|N(u) \cap N_3| \ge 2$, say $w,w' \in N(u) \cap N_3$, and recall that by Claim \ref{uinN2contactsforcedneighbor}, 
$u \in N_2 \cap Y$ contacts $N(v_2)$. 

First assume that $uv_1 \in E$. Since $(u,w,w',v_1,v_2,v_3,v_4,v_5)$ (with midpoint $u$) does not induce an $S_{1,1,5}$, we have $uv_3 \in E$. Analogously, 
if $uv_3 \in E$ and since $(u,w,w',v_3,v_2,v_1,v_6,v_5)$ (with midpoint $u$) does not induce an $S_{1,1,5}$, we have $uv_1 \in E$.

Now assume that $uv_1 \notin E$ and $uv_3 \notin E$. Then $uv \in E$ for $v \in N(v_2) \setminus \{v_1,v_3\}$. 
Since $(u,w,w',v,v_2,v_3,v_4,v_5)$ (with midpoint $u$) does not induce an $S_{1,1,5}$, we have $vv_4 \in E$. 
Analogously, since $(u,w,w',v,v_2,v_1,v_6,v_5)$ (with midpoint $u$) does not induce an $S_{1,1,5}$, we have $vv_6 \in E$. 
Thus, $(i)$ is shown, and analogously, $(ii)$ can be shown in the same way. 
Thus, Claim \ref{2DC6uinN2uv3oruv4} is shown. 
\qed

\begin{clai}\label{uinN2DneighbinN3twoinN3N4}
If for $u \in N_2$, the $D$-neighbor $r \in D \cap N_3$ has at least two neighbors in $N_3 \cup N_4$ then $uv_1 \notin E$, $uv_3 \notin E$ and $uv_4 \notin E$, 
$uv_6 \notin E$.  
\end{clai}

\noindent
{\bf Proof.} Recall that $v_1,v_3,v_5 \in X$ and $v_2,v_4,v_6 \in Y$.  
If $u \in N_2 \cap X$ then $uv_1 \notin E$, $uv_3 \notin E$, and analogously, if $u \in N_2 \cap Y$ then $uv_4 \notin E$, $uv_6 \notin E$. 

First assume that $u \in N_2 \cap Y$, and the $D$-neighbor of $u$ is $r \in D \cap N_3$ with neighbors $s,s' \in N_3 \cup N_4$. 
Since $(r,s,s',u,v_3,v_4,v_5,v_6)$ (with midpoint $r$) does not induce an $S_{1,1,5}$, we have $uv_3 \notin E$. 
Since $(r,s,s',u,v_1,v_2,v_3,v_4)$ (with midpoint $r$) does not induce an $S_{1,1,5}$, we have $uv_1 \notin E$. 

Analogously, if $u \in N_2 \cap X$ then $uv_4 \notin E$, $uv_6 \notin E$, and Claim \ref{uinN2DneighbinN3twoinN3N4} is shown.
\qed  

\begin{defi}\label{P4r2r3r4r5} 
Let $R=(r_2,r_3,r_4,r_5)$ be a $P_4$ in $G$ with $r_ir_{i+1} \in E$, $2 \le i \le 4$, endpoint $r_2 \in N_2$ and $r_3 \in N_3$, $r_4 \in N_3 \cup N_4$, 
$r_5 \in N_3 \cup N_4 \cup N_5$.
\end{defi}

\begin{clai}\label{P4r2r3r4r5r2contactsv3orv4} 
Let $R=(r_2,r_3,r_4,r_5)$ be such a $P_4$ as in Definition $\ref{P4r2r3r4r5}$.  
\begin{itemize}
\item[$(i)$] If $r_2 \in N_2 \cap Y$ then $r_2v_3 \in E$. 
\item[$(ii)$] If $r_2 \in N_2 \cap X$  then $r_2v_4 \in E$.
\end{itemize}
\end{clai}

\noindent
{\bf Proof.}
$(i)$: Let $r_2 \in N_2 \cap Y$. Clearly, by (\ref{(3)}), $|N(r_2) \cap N_3| \ge 2$, say $r_3,r'_3 \in N(r_2) \cap N_3$, and recall that by 
Claim \ref{uinN2contactsforcedneighbor}, $r_2 \in N_2 \cap Y$ contacts $N(v_2)$. 

Suppose to the contrary that $r_2v_3 \notin E$. If $r_2v_3 \notin E$ and $r_2v_1 \in E$ then $(r_2,r_3,r'_3,v_1,v_2,v_3,v_4,v_5)$ (with midpoint $r_2$) induce an $S_{1,1,5}$, which is a contradiction. Thus, $r_2v_1 \notin E$ and $r_2v \in E$ for $v \in N(v_2)$, $v \neq v_1,v_3$. But then 
$(v_2,v_1,v_3,v,r_2,r_3,r_4,r_5)$ (with midpoint $v_2$) induce an $S_{1,1,5}$, which is a contradiction. 
Thus, $(i)$ is shown, and analogously, $(ii)$ can be shown in the same way. 

Thus, Claim \ref{P4r2r3r4r5r2contactsv3orv4} is shown. 
\qed

\begin{coro}\label{2DP6C6P4r2inN2contactsv3orv4}
Let $R=(r_2,r_3,r_4,r_5)$ be such a $P_4$ as in Definition $\ref{P4r2r3r4r5}$. 
\begin{itemize}
\item[$(i)$] If $P=(v_1,\ldots,v_6)$ is a $2D-P_6$ with $v_2,v_5 \in D_{basis}$ then $r_2v_3 \in E$ or $r_2v_4 \in E$. 
\item[$(ii)$] If $C=(v_1,\ldots,v_6)$ is a $2D-C_6$ with $v_2,v_5 \in D_{basis}$ then $r_2v_1 \in E$ and $r_2v_3 \in E$ or $r_2v_4 \in E$ and $r_2v_6 \in E$.
\end{itemize}
\end{coro}

\noindent
{\bf Proof.}
Without loss of generality, assume that $r_2 \in N_2$ is white, i.e., $r_2 \in N_2 \cap Y$, and by Claim \ref{uinN2contactsforcedneighbor}, $r_2$ contacts $N(v_2)$. 

\medskip

\noindent
$(i)$: By Claim \ref{P4r2r3r4r5r2contactsv3orv4} $(i)$, $r_2v_3 \in E$. Analogously, if $r_2$ is black then by Claim \ref{P4r2r3r4r5r2contactsv3orv4} $(ii)$, $r_2v_4 \in E$. 

\medskip

\noindent
$(ii)$: Recall that by Claim \ref{2DC6uinN2uv3oruv4}, either $r_2v_1 \in E$ and $r_2v_3 \in E$ or $r_2v_1 \notin E$ and $r_2v_3 \notin E$.
By Claim \ref{P4r2r3r4r5r2contactsv3orv4} $(i)$, $r_2v_3 \in E$.
Thus, $r_2v_1 \in E$ and $r_2v_3 \in E$. 
Analogously, if $r_2$ is black then $r_2v_4 \in E$ and $r_2v_6 \in E$. 

\medskip

Thus, Corollary~\ref{2DP6C6P4r2inN2contactsv3orv4} is shown.
\qed

\begin{clai}\label{vinN3oneneighbinN3N4}
Let $R=(r_2,r_3,r_4,r_5)$ be such a $P_4$ as in Definition $\ref{P4r2r3r4r5}$. 
Then the $r_2$-neighbor $r_3 \in N_3$ has exactly one neighbor in $N_3 \cup N_4$, namely $r_4 \in N_3 \cup N_4$. 
\end{clai}

\noindent
{\bf Proof.}
Without loss of generality, assume that $r_2 \in N_2 \cap Y$, i.e., $r_2$ is white.  
Recall that by Claim \ref{P4r2r3r4r5r2contactsv3orv4} $(i)$, $r_2v_3 \in E$. 

Suppose to the contrary that $r_3$ has at least two neighbors $r_4,r'_4 \in N_3 \cup N_4$. 
Recall that $(v_1,\ldots,v_6)$ is a $2D-P_6$ or a $2D-C_6$ with $v_2,v_5 \in D$. But then $(r_3,r_4,r'_4,r_2,v_3,v_4,v_5,v_6)$ (with midpoint $r_3$) induce an $S_{1,1,5}$, which is a contradiction. 
Thus, $r_3 \in N_3$ has exactly one neighbor in $N_3 \cup N_4$. 

Analogously, if $r_2 \in N_2 \cap X$ then $r_3 \in N_3$ has exactly one neighbor in $N_3 \cup N_4$, and Claim \ref{vinN3oneneighbinN3N4} is shown. 
\qed

\begin{coro}\label{P4r2inN2contactsv3orv4r3notinD}
For every such $P_4$ $R=(r_2,r_3,r_4,r_5)$ as in Definition $\ref{P4r2r3r4r5}$, we have $r_3 \notin D$ and $r_4$ is $D_{basis}$-forced.
\end{coro}

\noindent
{\bf Proof.}
Let $R=(r_2,r_3,r_4,r_5)$ be such a $P_4$ as in Definition $\ref{P4r2r3r4r5}$. 
Without loss of generality, assume that $r_2 \in N_2 \cap Y$, i.e., $r_2$ is white. Recall that by Claim \ref{P4r2r3r4r5r2contactsv3orv4} $(i)$, $r_2v_3 \in E$. 

Suppose to the contrary that $r_3 \in D$. Then by the e.d.s.\ property, $r_5 \notin D$ and $r_5$ must have a $D$-neighbor 
$s \in D \cap (N_3 \cup N_4 \cup N_5 \cup N_6)$. 
But then $(v_3,v_2,v_4,r_2,r_3,r_4,r_5,s)$ (with midpoint $v_3$) induce an $S_{1,1,5}$, which is a contradiction.  
Thus, $r_3 \notin D$ and $r_3$ must have a $D$-neighbor in $D \cap (N_3 \cup N_4)$. 
Recall that by Claim \ref{vinN3oneneighbinN3N4}, $r_4$ is the only neighbor of $r_3$ in $N_3 \cup N_4$. Thus, $r_4$ is $D_{basis}$-forced, and  
 Corollary \ref{P4r2inN2contactsv3orv4r3notinD} is shown.
\qed

\medskip

Thus, for every such $P_4$ $R=(r_2,r_3,r_4,r_5)$, $D_{basis}$ can be updated, i.e., $D_{basis}:=D_{basis} \cup \{r_4\}$, and from now on, we have: 

\begin{coro}\label{noP4N2N3N4N5} 
There is no such $P_4$ $R=(r_2,r_3,r_4,r_5)$ as in Definition $\ref{P4r2r3r4r5}$. 
\end{coro}

\begin{clai}\label{N5emptyN4indep} 
$N_5=\emptyset$, $N_4$ is independent, and for every edge $(r_3,s_3)$ with $r_3,s_3 \in N_3$, $r_3 \cojoin N_4$ and $s_3 \cojoin N_4$.   
\end{clai}

\noindent
{\bf Proof.}
Recall that by Corollaries \ref{P4r2inN2contactsv3orv4r3notinD} and \ref{noP4N2N3N4N5} (and after updating $D_{basis}$), there is no such $P_4$ 
$R=(r_2,r_3,r_4,r_5)$ as in Definition $\ref{P4r2r3r4r5}$. Then $N_5=\emptyset$ and $N_4$ is independent.

Assume that there is an edge $r_3s_3 \in E$ with $r_3,s_3 \in N_3$ and $r_2r_3 \in E$ with $r_2 \in N_2$. 
If $s_3s_4 \in E$ for $s_4 \in N_4$ then $(r_2,r_3,s_3,s_4)$ induce a $P_4$, which is impossible by Corollary \ref{noP4N2N3N4N5}, i.e., $s_3 \cojoin N_4$ and analogously, $r_3 \cojoin N_4$.
Thus, Claim \ref{N5emptyN4indep} is shown. 
\qed 

\begin{clai}\label{P3inN3N4midpointN3} 
Let $(r,s,t)$ be a $P_3$ in $N_3 \cup N_4$ with midpoint $s \in N_3$. 
\begin{itemize}
\item[$(i)$] If $s \in N_3 \cap X$ then for $us \in E$ with $u \in N_2 \cap Y$, $uv_1 \notin E$ and $uv_3 \notin E$. 
Moreover, for $uv \in E$ with $v \in N(v_2)$, $v \join N(v_5)$. 
\item[$(ii)$] If $s \in N_3 \cap Y$ then for $us \in E$ with $u \in N_2 \cap X$, $uv_4 \notin E$ and $uv_6 \notin E$. 
Moreover, for $uv \in E$ with $v \in N(v_5)$, $v \join N(v_2)$. 
\end{itemize}
\end{clai}

\noindent
{\bf Proof.} $(i)$: Assume that there is a $P_3$ $(r,s,t)$ in $N_3 \cup N_4$ with midpoint $s \in N_3 \cap X$. Let $us \in E$ with $u \in N_2 \cap Y$. 
Recall that by Claim \ref{uinN2contactsforcedneighbor}, $u \in N_2 \cap Y$ contacts $N(v_2)$, and by Claim \ref{uinN2DneighbinN3twoinN3N4}, 
$uv_1 \notin E$ and $uv_3 \notin E$. Let $uv \in E$ with $v \in N(v_2)$, $v \neq v_1,v_3$.   

Since $(s,r,t,u,v,v_2,v_3,v_4)$ (with midpoint $s$) does not induce an $S_{1,1,5}$, we have $vv_4 \in E$, and analogously, since 
$(s,r,t,u,v,v_4,v_5,v_6)$ (with midpoint $s$) does not induce an $S_{1,1,5}$, we have $vv_6 \in E$. 
Moreover, if $v' \in N(v_5)$, $v' \neq v_4,v_6$, then, since $(s,r,t,u,v,v_4,v_5,v')$ (with midpoint $s$) does not induce an $S_{1,1,5}$, we have 
$vv' \in E$, and in general, $v \join N(v_5)$. 

\medskip

\noindent
$(ii)$: Analogously, the same can be done for a $P_3$ $(r,s,t)$ in $N_3 \cup N_4$ with midpoint $s \in N_3 \cap Y$.  

\medskip

Thus, Claim \ref{P3inN3N4midpointN3} is shown. 
\qed

\begin{clai}\label{P2+P3inN3N4midpointN3} 
Let $P=(r,s,t)$ be a $P_3$ in $N_3 \cup N_4$ with midpoint $s \in N_3$ and $P'=(s',t')$ be a $P_2$ with $s' \in N_3$, $t' \in N_3 \cup N_4$, such that either 
$s,s' \in X$ or $s,s' \in Y$, i.e., $s,s'$ have the same color, and $P,P'$ induce a $P_3+P_2$. Then $s$ and $s'$ do not have any distinct $N_2$-neighbors.
\end{clai}

\noindent
{\bf Proof.} 
Without loss of generality, assume that $P=(y_0,x_1,y_1)$ in $N_3 \cup N_4$ with midpoint $x_1 \in N_3 \cap X$ and $P'=(x_2,y_2)$ with $x_2 \in N_3 \cap X$ and 
$y_2 \in N_3 \cup N_4$ such that $P,P'$ induce a $P_3+P_2$. 

Suppose to the contrary that there are distinct $N_2$-neighbors of $x_1$ and $x_2$, say $u_1x_1 \in E$, $u_1x_2 \notin E$, and $u_2x_2 \in E$, $u_2x_1 \notin E$, 
for $u_1,u_2 \in N_2 \cap Y$. 
Recall that by Claim \ref{uinN2contactsforcedneighbor}, $u_1,u_2 \in N_2 \cap Y$ contact $N(v_2)$. 

If $u_1$ and $u_2$ have a common neighbor $v \in N(v_2)$ (possibly $v=v_1$ or $v=v_3$) then 
$(x_1,y_0,y_1,u_1,v,u_2,x_2,y_2)$ (with midpoint $x_1$) induce an $S_{1,1,5}$, which is a contradiction. Thus, $u_1$ and $u_2$ do not have any common neighbor 
$v \in N(v_2)$, say $u_1v \in E$ and $u_2v' \in E$ with $v,v' \in N(v_2)$, $v \neq v'$. 
But then $(x_1,y_0,y_1,u_1,v,v_2,v',u_2)$ (with midpoint $x_1$) induce an $S_{1,1,5}$, which is a contradiction. Thus, $x_1$ and $x_2$ do not have any distinct $N_2$-neighbors.

\medskip

Analogously, if there is a $P_3+P_2$ $(x_0,y_1,x_1)$ in $N_3 \cup N_4$ with midpoint $y_1 \in N_3 \cap Y$ and $(y_2,x_2)$ with $y_2 \in N_3 \cap Y$ and 
$x_2 \in N_3 \cup N_4$ then $y_1$ and $y_2$ do not have any distinct $N_2$-neighbors.

\medskip

Thus, Claim \ref{P2+P3inN3N4midpointN3} is shown. 
\qed 

\begin{clai}\label{P3inN3N4midpointDN3} 
If there is a $P_3$ $(r,s,t)$ in $N_3 \cup N_4$ with midpoint $s \in D \cap N_3 \cap X$ then $D \cap N_3 \cap X=\{s\}$.  
Analogously, if there is a $P_3$ $(r,s,t)$ in $N_3 \cup N_4$ with midpoint $s \in D \cap N_3 \cap Y$ then $D \cap N_3 \cap Y=\{s\}$.
\end{clai}

\noindent
{\bf Proof.}
Without loss of generality, assume that there is a $P_3$ $(y_0,x_1,y_1)$ in $N_3 \cup N_4$ with midpoint $x_1 \in D \cap N_3 \cap X$. 

Suppose to the contrary that there is a second $D$-neighbor $x_2 \in D \cap N_3 \cap X$. Clearly, by (\ref{(4)}), $x_2$ has a neighbor $y_2 \in N_3 \cup N_4$ (else $x_2$ is $D_{basis}$-forced), and thus, $(y_0,x_1,y_1)$, $(x_2,y_2)$ induce a $P_3+P_2$. Let $u_1x_1 \in E$ with $u_1 \in N_2$ and $u_2x_2 \in E$ with $u_2 \in N_2$. By the e.d.s.\ property, $x_1$ and $x_2$ do not have any common $N_2$-neighbor, i.e., $u_1 \neq u_2$,  
but then by Claim \ref{P2+P3inN3N4midpointN3}, it is a contradiction.  

\medskip

Analogously, if there is a $P_3$ $(r,s,t)$ in $N_3 \cup N_4$ with midpoint $s \in D \cap N_3 \cap Y$ then $D \cap N_3 \cap Y=\{s\}$.   

\medskip

Thus, Claim \ref{P3inN3N4midpointDN3} is shown. 
\qed

\begin{coro}\label{coroP3inN3N4midpointDN3} 
There is at most one $P_3$ $(r,s,t)$ in $N_3 \cup N_4$ with midpoint $s \in D \cap N_3 \cap X$ and at most one $P_3$ $(r',s',t')$ in $N_3 \cup N_4$ with midpoint 
$s' \in D \cap N_3 \cap Y$.
\end{coro}

Then $D_{basis}$ can be updated in polynomial time, i.e., $D_{basis}:=D_{basis} \cup \{s,s'\}$, and we have: 

\begin{coro}\label{coro:noP3midpointDN3} 
There is no such $P_3$-midpoint $s \in D \cap N_3 \cap X$ or $s' \in D \cap N_3 \cap Y$ with $P_3$ in $N_3 \cup N_4$.  
\end{coro}

Then for every such $P_3$ in $N_3 \cup N_4$, it leads to a $P_4$ $(x_1,y_1,x_2,y_2)$ in $N_3 \cup N_4$ with $x_1,y_2 \in D$ (recall that by Claim 
\ref{N5emptyN4indep}, $N_5=\emptyset$, $N_4$ is independent and every edge in $N_3$ does not contact $N_4$). 

\begin{clai}\label{oneP4inN3} 
If there is a $P_4$ $(x_1,y_1,x_2,y_2)$ in $N_3$ with $x_1,y_2 \in D$ then $N(x_1) \cap (N_3 \cup N_4)=\{y_1\}$, $N(y_2) \cap (N_3 \cup N_4)=\{x_2\}$,
and $D \cap N_3 = \{x_1,y_2\}$.  
\end{clai}

\noindent
{\bf Proof.}
Assume that there is a $P_4$ $(x_1,y_1,x_2,y_2)$ in $N_3$ with $x_1,y_2 \in D$. By Corollary~\ref{coro:noP3midpointDN3}, $x_1$ and $y_2$ are no such $P_3$-midpoints  with $P_3$ in $N_3 \cup N_4$, i.e., $N(x_1) \cap (N_3 \cup N_4)=\{y_1\}$ and $N(y_2) \cap (N_3 \cup N_4)=\{x_2\}$. 

Suppose to the contrary that there is another $D$-vertex in $N_3$. Assume without loss of generality that $x_3 \in D \cap N_3$, and by (\ref{(4)}), let 
$x_3y_3 \in E$ for $y_3 \in N_3 \cup N_4$ (else $x_3$ is $D_{basis}$-forced). 
Let $u_1x_1 \in E$ for $u_1 \in N_2$ and $u_3x_3 \in E$ for $u_3 \in N_2$. Clearly, by the e.d.s.\ property, $u_1 \neq u_3$, i.e., 
$u_1x_3 \notin E$ and $u_3x_1 \notin E$, $x_3y_1 \notin E$, and $x_3y_2 \notin E$. 

Since by Corollary~\ref{noP4N2N3N4N5}, $(u_1,x_1,y_1,x_2)$ does not induce a $P_4$, we have $u_1x_2 \in E$. 
Since by the e.d.s.\ property, $u_3x_1 \notin E$, and by Corollary~\ref{noP4N2N3N4N5}, $(u_3,x_2,y_1,x_1)$ does not induce a $P_4$, we have $u_3x_2 \notin E$. 
Moreover, since by Corollary~\ref{noP4N2N3N4N5}, $(u_1,x_2,y_3,x_3)$ does not induce a $P_4$, we have $y_3x_2 \notin E$, i.e., $(y_1,x_2,y_2)$, $(x_3,y_3)$ induce a $P_3+P_2$ in $G$, and $x_2,x_3$ do not have any common neighbor in $N_2$, which is a contradiction by Claim \ref{P2+P3inN3N4midpointN3}. 
Thus, there is no other $D$-vertex $x_3 \in D \cap X \cap N_3$.  

\medskip

Analogously, there is no other $D$-vertex $y_3 \in D \cap Y \cap N_3$, and Claim \ref{oneP4inN3} is shown. 
\qed

\medskip

Then $D_{basis}$ can be updated, i.e., $D_{basis}:=D_{basis} \cup \{x_1,y_2\}$. 

\begin{coro}\label{oneP4inN3N4} 
If there is a $P_4$ $(x_1,y_1,x_2,y_2)$ in $N_3 \cup N_4$ with $x_1,x_2 \in N_3$, $y_1,y_2 \in N_4$, and $x_1 \in D$ then $D \cap X \cap N_3 = \{x_1\}$.  
Analogously, if there is a $P_4$ $(y_1,x_1,y_2,x_2)$ in $N_3 \cup N_4$ with $y_1,y_2 \in N_3$, $x_1,x_2 \in N_4$, and $y_1 \in D$ then 
$D \cap Y \cap N_3 = \{y_1\}$. 
\end{coro}

\noindent
{\bf Proof.}
Assume without loss of generality that there is a $P_4$ $(x_1,y_1,x_2,y_2)$ in $N_3 \cup N_4$ with $x_1,x_2 \in N_3$, $y_1,y_2 \in N_4$, and $x_1 \in D$.
By the e.d.s.\ property, $x_2 \notin D$ and $x_2$ must have a $D$-neighbor in $N_3 \cup N_4$, and by Claim \ref{N5emptyN4indep}, $x_2$ does not have any neighbor in $N_3$ but $x_2$ must have a $D$-neighbor in $N_4$. If $|N(x_2) \cap N_4| \ge 3$, say $y_1,y_2,y'_2 \in N(x_2) \cap N_4$, and without loss of generality, $y_2 \in D$ then $y'_2$ must have a $D$-neighbor in $N_3$ (possibly $x_1y'_2 \in E$).    
Since $x_1$ has a neighbor $u_1 \in N_2$, i.e., $u_1x_1 \in E$, then by Corollary \ref{noP4N2N3N4N5}, $u_1x_2 \in E$. If $y'_2x_1 \notin E$, i.e., $y'_2x_3 \in E$ with $x_3 \in D$ then $u_1x_3 \notin E$ but then $(u_1,x_2,y'_2,x_3)$ induce a $P_4$ which is a contradiction by Corollary \ref{noP4N2N3N4N5}.  
Thus, $x_1y'_2 \in E$ but then $(y_1,x_1,y'_2)$ induce a $P_3$ with midpoint $x_1 \in D \cap N_3$, which is impossible by Corollary \ref{coro:noP3midpointDN3}. 
Thus, $N(x_2) \cap N_4=\{y_1,y_2\}$. 

\medskip

Suppose to the contrary that there is a second $D$-vertex in $N_3 \cap X$, say $x_3 \in D \cap N_3 \cap X$. Clearly, by the e.d.s.\ property, 
$x_1$ and $x_3$ do not have any common neighbor in $N_2$, say $u_1x_1 \in E$, $u_3x_3 \in E$, $u_1,u_3 \in N_2$, $u_1x_3 \notin E$, $u_3x_1 \notin E$, as well as 
$x_3y_1 \notin E$ and $x_3y_2 \notin E$. Recall that $u_3x_3 \in E$ with $u_3 \in N_2$, and by Corollary \ref{noP4N2N3N4N5}, $u_3x_2 \notin E$ (else there is a $P_4$ $(u_3,x_2,y_1,x_1)$ which is impossible).
Let $x_3y_3 \in E$ with $y_3 \in N_3 \cup N_4$ (else $x_3$ is $D_{basis}$-forced). Clearly, by the e.d.s.\ property, $y_3x_1 \notin E$. 
Then by Corollary \ref{noP4N2N3N4N5}, $y_3x_2 \notin E$ (else there is a $P_4$ $(u_3,x_3,y_3,x_2)$ which is impossible), and thus, 
 $(y_1,x_2,y_2)$ and $(x_3,y_3)$ induce a $P_3+P_2$ in $G$.  

Recall that $u_1x_1 \in E$, $u_3x_3 \in E$ with $u_1,u_3 \in N_2$, and by the e.d.s.\ property, $u_1 \neq u_3$, $u_1x_3 \notin E$, $u_3x_1 \notin E$. 
Thus, $x_1$ and $x_3$ do not have any common $N_2$-neighbor, which is a contradiction by Claim \ref{P2+P3inN3N4midpointN3}. 
Thus, Corollary \ref{oneP4inN3N4} is shown.
\qed

\medskip

Analogously, $D_{basis}$ can be updated, i.e., $D_{basis}:=D_{basis} \cup \{x_1,y_2\}$. 
Now assume that there is no such $P_4$ in $N_3 \cup N_4$ and there is no such $P_3$ in $N_3 \cup N_4$ with midpoint in $N_3$. 
Then for every $P_2$ $(x,y)$ in $N_3$, either $x \in D$ or $y \in D$ (else there is no such e.d.s.\ in $G$ with $D_{basis}$). 

\begin{clai}\label{3P2inN3} 
If there are two $P_2$'s $(x_1,y_1)$, $(x_2,y_2)$ in $N_3$ with $u_1x_1 \in E$ and $u_1x_2 \in E$ for $u_1 \in N_2$ as well as $u_2y_1 \in E$ and $u_2y_2 \in E$ for $u_2 \in N_2$ then either $x_1,y_2 \in D$ or $x_2,y_1 \in D$. Moreover, there are no three such $P_2$'s with common $N_2$-neighbors.  
\end{clai}

\noindent
{\bf Proof.}
Without loss of generality, assume that for $P_2$ $(x_1,y_1)$ in $N_3$, $x_1 \in D$. Then by the e.d.s.\ property, $x_2 \notin D$ and $y_2 \in D$. 
Analogously, if $y_1 \in D$ then $y_2 \notin D$ and $x_2 \in D$. 

If $x_1,y_2 \in D$ or $x_2,y_1 \in D$ and there is a third $P_2$ $(x_3,y_3)$ with common $N_2$-neighbors $u_1,u_2$ then by the e.d.s.\ property, 
$x_3,y_3 \notin D$ and $x_3,y_3$ do not have any $D$-neighbor, which is impossible, i.e., there is no such e.d.s.
Thus, Claim \ref{3P2inN3} is shown. 
\qed

\begin{coro}\label{uinN2contacts3P2inN3} 
If $u \in N_2$ contacts three $P_2$'s $(x_1,y_1)$, $(x_2,y_2)$, $(x_3,y_3)$ in $N_3$ with $ux_i \in E$, $1 \le i \le 3$, and $u'y_1 \in E$, $u'y_2 \in E$ then 
$x_3 \notin D$ and $y_3 \in D$ is $D_{basis}$-forced. Moreover, if $u'y_i \in E$, $1 \le i \le 3$, then there is no such e.d.s. 
\end{coro}

\noindent
{\bf Proof.}
Suppose to the contrary that $x_3 \in D$. Then $x_1,x_2 \notin D$ and $y_1,y_2 \in D$, which is a contradiction by the e.d.s.\ property, since $u'y_1 \in E$, $u'y_2 \in E$. Thus, $x_3 \notin D$ and $y_3 \in D$ is $D_{basis}$-forced.

If $u'y_i \in E$, $1 \le i \le 3$, then assume that $x_1 \in D$, i.e., $x_2,x_3 \notin D$ and $y_2,y_3 \in D$. Recall that by Claim \ref{3P2inN3}, there are no three such $P_2$'s with common $N_2$-neighbors. Then there is no such e.d.s.
\qed 

\medskip

By the way, if there are three $P_2$'s $(x_1,y_1)$, $(x_2,y_2)$, $(x_3,y_3)$ in $N_3$ with $u_1x_i \in E$, $1 \le i \le 3$, and $x_1 \in D \cap N_3$ as well as $y_2,y_3$ have a common $N_2$-neighbor then there is no such e.d.s. (recall that if $x_1 \in D \cap N_3$ then $x_2,x_3 \notin D$ and $y_2,y_3 \in D$).

\begin{clai}\label{noP6inN2N3} 
There is no $P_6$ $(u_1,w_1,u_2,w_2,u_3,w_3)$ in $N_2 \cup N_3$ with $u_i \in N_2$ and $w_i \in N_3$, $1 \le i \le 3$. 
\end{clai}

\noindent
{\bf Proof.}
Without loss of generality, suppose to the contrary that there is a $P_6$ $(u_1,x_1,u_2,x_2,u_3,x_3)$ in $N_2 \cup N_3$ with $u_i \in N_2 \cap Y$ and 
$x_i \in N_3 \cap X$, $1 \le i \le 3$.
Clearly, $x_1$ must have a neighbor $y_1 \in N_3 \cup N_4$ (else $x_1$ is $D_{basis}$-forced), and analogously, $x_3$ must have a neighbor $y_3 \in N_3 \cup N_4$. 
Since by Corollary \ref{noP4N2N3N4N5}, $y_1x_2 \notin E$, $y_1x_3 \notin E$, we have $y_1 \neq y_3$, i.e., $x_3y_3 \in E$ with $y_3 \in N_3 \cup N_4$, and again by 
Corollary~\ref{noP4N2N3N4N5}, $x_1y_3 \notin E$, $x_2y_3 \notin E$.
But then $(x_1,u_1,y_1,u_2,x_2,u_3,x_3,y_3)$ (with midpoint $x_1$) induce an $S_{1,1,5}$, which is a contradiction. Thus, there is no such 
$P_6$ $(u_1,x_1,u_2,x_2,u_3,x_3)$ in $N_2 \cup N_3$ with $u_i \in N_2 \cap Y$ and $x_i \in N_3 \cap X$, $1 \le i \le 3$.

Analogously, there is no such 
$P_6$ $(u_1,y_1,u_2,y_2,u_3,y_3)$ in $N_2 \cup N_3$ with $u_i \in N_2 \cap X$ and $y_i \in N_3 \cap Y$, $1 \le i \le 3$.
Thus, Claim \ref{noP6inN2N3} is shown.
\qed  

\begin{coro}\label{P5inN2N3notinD} 
If there is a $P_5$ $(u_1,w_1,u_2,w_2,u_3)$ in $N_2 \cup N_3$ with $u_i \in N_2$, $1 \le i \le 3$, and $w_1,w_2 \in N_3$ then $w_1,w_2 \notin D$, and there is a 
common $D$-neighbor $w \in D \cap N_3$ with $wu_i$, $1 \le i \le 3$.  
\end{coro}

\noindent
{\bf Proof.}
Without loss of generality, assume that $(u_1,x_1,u_2,x_2,u_3)$ induce a $P_5$ with $u_i \in N_2 \cap Y$, $1 \le i \le 3$, and $x_i \in N_3 \cap X$, $1 \le i \le 2$. 

\medskip

Suppose to the contrary that $x_1 \in D$ or $x_2 \in D$; without loss of generality, assume $x_1 \in D$. Then $x_2 \notin D$ and $u_3$ must have a $D$-neighbor $x_3 \in D \cap N_3$, i.e., $u_3x_3 \in E$. 
By the e.d.s.\ property, $x_3u_1 \notin E$ and $x_3u_2 \notin E$ but then $(u_1,x_1,u_2,x_2,u_3,x_3)$ induce a $P_6$, which is impossible by Claim \ref{noP6inN2N3}.
Thus, $x_1,x_2 \notin D$. Let $x'_i \in D \cap N_3 \cap X$, $1 \le i \le 3$, be $D$-neighbors of $u_i$. 

If for $x'_3u_3 \in E$, we have $x'_3u_1 \notin E$ and $x'_3u_2 \notin E$ then $(u_1,x_1,u_2,x_2,u_3,x'_3)$ induce a $P_6$ which is impossible by 
Claim \ref{noP6inN2N3}. Analogously, if for $x'_1u_1 \in E$, we have $x'_1u_2 \notin E$ and $x'_1u_3 \notin E$ then $(u_3,x_2,u_2,x_1,u_1,x'_1)$ induce a $P_6$ which is impossible by Claim \ref{noP6inN2N3}. 

If for $x'_3u_3 \in E$, we have $x'_3u_1 \notin E$ and $x'_3u_2 \in E$ then for the $D$-neighbor $x'_1 \in D \cap N_3$ of $u_1$, by the e.d.s.\ property, 
$x'_1u_2 \notin E$ and $x'_1u_3 \notin E$ but then, $(u_3,x'_3,u_2,x_1,u_1,x'_1)$ induce a $P_6$ which is impossible by 
Claim \ref{noP6inN2N3}. 
Analogously, if for $x'_3u_3 \in E$, we have $x'_3u_1 \in E$ and $x'_3u_2 \notin E$ then $(u_3,x'_3,u_1,x_1,u_2,x'_2)$ induce a $P_6$ which is impossible by Claim \ref{noP6inN2N3}. Thus, there is a common $D$-neighbor $x \in D \cap N_3 \cap X$ for $u_i$, $1 \le i \le 3$. 

Analogously, if $(u_1,y_1,u_2,y_2,u_3)$ induce a $P_5$ with $u_i \in N_2 \cap X$, $1 \le i \le 3$, and $y_i \in N_3 \cap Y$, $1 \le i \le 2$, then $y_1,y_2 \notin D$ and there is a common $D$-neighbor $y \in D \cap N_3 \cap Y$ for $u_i$, $1 \le i \le 3$. 

\medskip

Thus, Corollary \ref{P5inN2N3notinD} is shown.    
\qed   

\medskip

Let $Q$ be a component in $G[N_2 \cup N_3 \cup N_4]$, say $V(Q)=N^Q_2 \cup N^Q_3 \cup N^Q_4$. Recall that by $(\ref{(1)})$, $D \cap N_2=\emptyset$, i.e., 
$D \cap N^Q_2=\emptyset$. 
If there is an e.d.s.\ $D$ in $Q$ then for a component $Q'$ with $N^{Q'}_2 =(N^Q_2 \cap X) \join (N^Q_2 \cap Y)$ and $N^{Q'}_3=N^Q_3$, $N^{Q'}_4=N^Q_4$, 
 $Q'$ has the same e.d.s.\ $D$ in $Q'$ as in $Q$. Thus assume: 
\begin{equation}\label{(N2XY)}
(N^Q_2 \cap X) \join (N^Q_2 \cap Y).
\end{equation}

\begin{coro}\label{dist2or4inN3XorN3Y} 
For every $w,w' \in V(Q) \cap N_3 \cap X$ or $w,w' \in V(Q) \cap N_3 \cap Y$, $w \neq w'$, $dist_Q(w,w') =2$ or $dist_Q(w,w') =4$.  
\end{coro}

\noindent
{\bf Proof.}
Without loss of generality, assume that $x_1,x_2 \in V(Q) \cap X$. If $x_1$ and $x_2$ have a common neighbor in $N_2 \cup N_4$ then $dist_Q(x_1,x_2)=2$.
Now assume that $x_1$ and $x_2$ do not have any common neighbor in $N_2 \cup N_4$. Let $u_ix_i \in E$ with $u_i \in N_2 \cap Y$, $i \in \{1,2\}$, $u_1 \neq u_2$. 
Moreover, since there is no such $P_3$ in $N_3$, $x_1$ and $x_2$ do not have any common neighbor in $N_3$. 
By (\ref{(N2XY)}), $u_1,u_2$ have a common neighbor $x \in N_2$, and then $dist_Q(x_1,x_2)=4$.  
\qed

\medskip

Now the set $N_2 \cup N_3 \cup N_4$ can be partitioned into two subsets, namely, $H_1$ and $H_2$ such that
$H_1 := [(N_2 \cup N_4) \cap X] \cup [N_3 \cap Y]$ and $H_2 := [(N_2 \cup N_4) \cap Y] \cup [N_3 \cap X]$.

\subsection{When $H_1 = \emptyset$ or $H_2 = \emptyset$}\label{H1emptyorH2empty}

Assume without loss of generality that $H_1 = \emptyset$. Then $N_2 \cup N_4 \subset Y$ and $N_3 \subset X$, i.e., $N_i$ are independent, $2 \le i \le 4$. 
Recall that every $x \in N_3$ does not have a neighbor in $N_3$ (since $N_3$ is independent) but $x$ must have a neighbor in $N_4$ (else $x$ is $D_{basis}$-forced).

The e.d.s.\ problem can be solved independently for every component $Q$ in $G[N_2 \cup N_3 \cup N_4]$. 
If for $Q$, $|D \cap N_3| \le 2$ then the e.d.s.\ problem can be easily solved in polynomial time. Now assume that $|D \cap N_3|=k$, $k \ge 3$. 

For every $P_3$ $(u_i,x_i,y_i)$ with $x_i \in D \cap N_3$, $u_i \in N_2$ and $y_i \in N_4$, $1 \le i \le k$, $(u_1,x_1,y_1),\ldots,(u_k,x_k,y_k)$ induce a $kP_3$ in $Q$ . 
If $xy_i \in E$ for $x \in N_3$ then, since by Corollary \ref{noP4N2N3N4N5}, there is no such $P_4$, we have $xu_i \in E$.    
Moreover, if $xu_i \in E$ and $xu_j \in E$ with $u_ix_i \in E$, $x_i \in D$, and $u_jx_j \in E$, $x_j \in D$, $i \neq j$, then $xy_i \notin E$ and $xy_j \notin E$ since by Corollary \ref{noP4N2N3N4N5}, there is no such $P_4$ in $Q$. 

\begin{clai}\label{DcapN3atleast3xjoinN2}
If in $Q$, there is a $3P_3$ $(u_1,x_1,y_1)$, $(u_2,x_2,y_2)$, $(u_3,x_3,y_3)$,  with $u_i \in N_2$, $x_i \in N_3$, $y_i \in N_4$, $1 \le i \le 3$, and $x \in N_3$ contacts $u_1,u_2$ or $u_2,u_3$ then $x \join V(Q) \cap N_2$. 
\end{clai}

\noindent
{\bf Proof.} Assume without loss of generality that $x \in N_3$ contacts $u_1,u_2$, i.e., $xu_1 \in E$, $xu_2 \in E$. 

Suppose to the contrary that $x$ does not join $V(Q) \cap N_2$, i.e., $xu_3 \notin E$, say $x'u_2 \in E$, $x'u_3 \in E$ and $x'u_1 \notin E$. 
But then $(u_1,x,u_2,x',u_3,x_3)$ induce a $P_6$ in $Q$, which is impossible by Claim \ref{noP6inN2N3}.
Thus, by Corollary \ref{P5inN2N3notinD}, $x \join N_2$, and Claim \ref{DcapN3atleast3xjoinN2} is shown. 
\qed
 
\medskip

Then for every component $Q$, either $x \in D$ and for all such $P_3$ $(u_i,x_i,y_i)$, $y_i \in D$, or $x \notin D$, $y \in D \cap N_4$ with $xy \in E$, and 
$x_i \in D$, $1 \le i \le k$.  
Thus, for every component in $G[N_2 \cup N_3 \cup N_4]$, the e.d.s.\ problem can be done in polynomial time.

\subsection{When $H_1 \neq \emptyset$ and $H_2 \neq \emptyset$}\label{H1H2nonempty}

Recall that by Claim \ref{N5emptyN4indep}, $N_5=\emptyset$, $N_4$ is independent, and every edge in $N_3$ does not contact $N_4$,  
and by Corollary \ref{dist2or4inN3XorN3Y}, in every component $Q$, $dist_Q(x,x') =2$ or $dist_Q(x,x') =4$ for every $x,x' \in N_3 \cap X$, and $dist_Q(y,y') =2$ or $dist_Q(y,y') =4$ for every $y,y' \in N_3 \cap Y$.

\medskip

If there is no $P_2$ in $N_3$, i.e., there is no contact between components in $H_1 \cap (N_3 \cup N_4)$ and $H_2 \cap (N_3 \cup N_4)$ then the e.d.s.\ problem can be done in polynomial time as in section \ref{H1emptyorH2empty}. 
Now assume that there exist $P_2$'s in $N_3$, i.e., there are such contacts between components in $H_1 \cap (N_3 \cup N_4)$ and $H_2 \cap (N_3 \cup N_4)$. 
By (\ref{(N2XY)}), $(N^Q_2 \cap X) \join (N^Q_2 \cap Y)$. 

\medskip

Recall Claims \ref{P3inN3N4midpointN3}, \ref{P2+P3inN3N4midpointN3}, \ref{P3inN3N4midpointDN3} and Corollaries \ref{coroP3inN3N4midpointDN3}, 
\ref{coro:noP3midpointDN3} as well as Claims \ref{oneP4inN3}, \ref{3P2inN3}, \ref{noP6inN2N3} and Corollaries \ref{oneP4inN3N4}, \ref{uinN2contacts3P2inN3}, \ref{P5inN2N3notinD}, \ref{dist2or4inN3XorN3Y}.

Then there is no $P_4$ and no $P_3$ in $Q[N_3]$, and for every $P_2$ in $Q[N_3]$, say $P=(x,y)$, either $x \in D$ or $y \in D$. 
If there are at most two such $P_2$'s $(x_1,y_1)$, $(x_2,y_2)$ in $Q[N_3]$ then either $x_1 \in D$ or $y_1 \in D$ as well as $x_2 \in D$ or $y_2 \in D$
and it can be done in polynomial time. 

\begin{clai}\label{3P2inN3notwocommonN2}
If there is a $3P_2$ $(x_1,y_1)$, $(x_2,y_2)$, $(x_3,y_3)$ in $N_3$ then there are no common $N_2$-neighbors $ux_i \in E$ and $u'y_i \in E$, $1 \le i \le 3$.  
\end{clai}

\noindent
{\bf Proof.}
Suppose to the contrary that there are such common neighbors $u,u' \in N_2$, with $ux_i \in E$, $u'y_i \in E$, $1 \le i \le 3$.
Recall that for every $P_2$ $(x_i,y_i)$, either $x_i \in D$ or $y_i \in D$. Without loss of generality, assume that $x_1 \in D$. Then by the e.d.s.\ property, $x_2,x_3 \notin D$ and $y_2,y_3 \in D$. But then $u'$ has two $D$-neighbors, which is a contradiction by the e.d.s.\ property. 
Thus, there are no common $N_2$-neighbors $ux_i \in E$, $u'y_i \in E$, $1 \le i \le 3$.
\qed

\begin{clai}\label{3P2inN3commonN2}
If for a $3P_2$ $(x_1,y_1)$, $(x_2,y_2)$, $(x_3,y_3)$ in $N_3$, there exist $u,u' \in N_2$ with $ux_i \in E$, $1 \le i \le 3$, and $u'y_i \in E$, $1 \le i \le 2$, or 
$uy_i \in E$, $1 \le i \le 3$, and $u'x_i \in E$, $1 \le i \le 2$, then $x_3 \notin D$ and $y_3 \in D$ is $D_{basis}$-forced or $y_3 \notin D$ and $x_3 \in D$ is $D_{basis}$-forced.  
\end{clai}

\noindent
{\bf Proof.}
Without loss of generality, assume that $ux_i \in E$, $1 \le i \le 3$, and $u'y_i \in E$, $1 \le i \le 2$. 
 
Suppose to the contrary that $x_3 \in D$. Then $x_1,x_2 \notin D$ and $y_1,y_2 \in D$. But then $u'$ has two $D$-neighbors, which is a contradiction by the e.d.s.\ property. Thus, $x_3 \notin D$ and $y_3 \in D$ is $D_{basis}$-forced. 

Analogously, if $uy_i \in E$, $1 \le i \le 3$, and $u'x_i \in E$, $1 \le i \le 2$, then $y_3 \notin D$ and $x_3 \in D$ is $D_{basis}$-forced.  
\qed

\begin{coro}\label{coro4P2inN3commonN2}
If for a $4P_2$ $(x_i,y_i)$, $1 \le i \le 4$, in $N_3$, there exist $u,u',u'' \in N_2$ with $ux_i \in E$, $1 \le i \le 4$, $u'y_i \in E$, $1 \le i \le 2$,  
$u''y_i \in E$, $3 \le i \le 4$, or $uy_i \in E$, $1 \le i \le 4$, $u'x_i \in E$, $1 \le i \le 2$, $u''x_i \in E$, $3 \le i \le 4$,
then there is no such e.d.s.\ in $Q$.
\end{coro}

\noindent
{\bf Proof.}
Without loss of generality, assume that $ux_i \in E$, $1 \le i \le 4$, $u'y_i \in E$, $1 \le i \le 2$, $u''y_i \in E$, $3 \le i \le 4$, and without loss of generality,
say $x_1 \in D$. Then $x_3,x_4 \notin D$ and $y_3,y_4 \in D$ but then $u''$ has two $D$-neighbors, which is a contradiction by the e.d.s.\ property. 
\qed

\begin{clai}\label{noP8inQ}
Every $Q$ is $P_8$-free.   
\end{clai}

\noindent
{\bf Proof.}
Recall that by (\ref{(N2XY)}), for every $u_1,u_2 \in V(Q) \cap N_2$, $dist_Q(u_1,u_2) \le 2$. Moreover, $N_5=\emptyset$, $N_4$ is independent, there is no $P_3$ in $N_3$, and for every $P_2$ $(x,y)$ with $u_1x \in E$ and $u_2y \in E$, $u_1u_2 \in E$. If for $x \in N_3$, $ux \in E$ with $u \in N_2$ and $xy \in E$, $x'y \in E$ with $y \in N_4$, then, since there is no such $P_4$ $(u,x,y,x')$, we have $ux' \in E$. 
Recall that by Claim \ref{noP6inN2N3}, there is no $P_6$ $(u_1,w_1,u_2,w_2,u_3,w_3)$ in $N_2 \cup N_3$ with $u_i \in N_2$ and $w_i \in N_3$, $1 \le i \le 3$. 
If there is a $P_7$ $P=(y_1,x_1,u_1,u,u_2,x_2,y_2)$ with $u_1,u,u_2 \in N_2$, $x_1,x_2 \in N_3$, $y_1,y_2 \in N_3 \cup N_4$ then there is no such $P_8$ with $P$.
Analogously, if there is a $P_7$ $P=(y_1,x_1,u_1,x_2,u_2,x_3,y_3)$ with $u_1,u_2 \in N_2$, $x_1,x_2,x_3 \in N_3$, $y_1,y_3 \in N_3 \cup N_4$ then there is no such $P_8$ with $P$. Thus, Claim \ref{noP8inQ} is shown.
\qed

\medskip

In \cite{BraMos2020P8}, we showed that ED is solvable in polynomial time for $P_8$-free bipartite graphs.
Thus, for every such $Q$, the e.d.s.\ problem can be solved independently in polynomial time. 

\section{When there is no $2D-P_6$ and no $2D-C_6$ in $G$}\label{no2DP6C6}

In this section, assume that there is no such $2D-P_6$ or $2D-C_6$ $(w_1,\ldots,w_6)$ with $w_2,w_5 \in D$. Recall that $G$ is connected.

If $|D|=1$ in $G$ then the e.d.s.\ problem is solvable in polynomial time. Now assume that $|D| \ge 2$ in $G$. 

\begin{clai}\label{P73th5thnotinD}
If $P=(v_1,\ldots,v_7)$ induce a $P_7$ in $G$ then $v_3 \notin D$ and $v_5 \notin D$. 
\end{clai}

\noindent
{\bf Proof.}
Without loss of generality, assume that $P=(x_1,y_1,x_2,y_2,x_3,y_3,x_4)$ induce a $P_7$ in $G$. Suppose to the contrary that $x_2 \in D$ or $x_3 \in D$, say without loss of generality, $x_2 \in D$. Then by the e.d.s.\ property, $x_1,y_1,y_2,x_3 \notin D$ and $x_1$ as well as $x_3$ must have a $D$-neighbor. 

Let $y \in D$ with $x_1y \in E$. Clearly, $y \neq y_1$ since $y_1 \notin D$. If $yx_3 \in E$ then $(y,x_1,y_1,x_2,y_2,x_3)$ induce a $2D-C_6$ which is impossible in this section.  
Thus, $yx_3 \notin E$. Then $x_3$ must have a $D$-neighbor $y' \in D$. If $y'=y_3$ then $(y_1,x_2,y_2,x_3,y_3,x_4)$ induce a $2D-P_6$ which is impossible in this section. Thus, $y_3 \notin D$ and $y' \neq y_3$. But then $(x_3,y_3,y',y_2,x_2,y_1,x_1,y)$ (with midpoint $x_3$) induce an $S_{1,1,5}$, 
which is a contradiction. 
Thus, $x_2 \notin D$ and analogously, $x_3 \notin D$, i.e., the third and fifth vertex $x_2,x_3$ of every $P_7$ $P=(x_1,y_1,x_2,y_2,x_3,y_3,x_4)$ is not in $D$. 

\medskip

Analogously, if $P=(y_1,x_1,\ldots,x_3,y_4)$ induce a $P_7$ then $y_2 \notin D$ and $y_3 \notin D$, and Claim \ref{P73th5thnotinD} is shown.
\qed

\begin{coro}\label{P8P4midnotinD}
If $P=(v_1,v_2,v_3,v_4,v_5,v_6,v_7,v_8)$ induce a $P_8$ in $G$ then $v_3,v_4,v_5,v_6 \notin D$.
\end{coro}

\begin{lemma}\label{P8secondinD}
If $P=(v_1,v_2,\ldots,v_7,v_8)$ induce a $P_8$ in $G$ then $v_2 \in D$ and $v_7 \in D$.
\end{lemma}

\noindent
{\bf Proof.}
Assume that $P=(x_1,y_1,x_2,y_2,x_3,y_3,x_4,y_4)$ induce a $P_8$ in $G$. 
Recall that by Corollary \ref{P8P4midnotinD}, $x_2,y_2,x_3,y_3 \notin D$. 

Suppose to the contrary that $y_1 \notin D$ or $x_4 \notin D$.

\medskip

\noindent
{\bf Case 1.} $y_1 \notin D$ and $x_4 \notin D$: 

\medskip

\noindent
Then $x_2$ must have a $D$-neighbor $y \in D$, and $y_3$ must have a $D$-neighbor $x \in D$.

Since $(x_2,y,y_1,y_2,x_3,y_3,x_4,y_4)$ (with midpoint $x_2$) does not induce an $S_{1,1,5}$, we have $yx_3 \in E$ or $yx_4 \in E$. 
Analogously, since $(y_3,x,x_4,x_3,y_2,x_2,y_1,x_1)$ (with midpoint $y_3$) does not induce an $S_{1,1,5}$, we have $xy_1 \in E$ or $xy_2 \in E$.   

\medskip

\noindent
{\bf Case 1.1} $xy_1 \in E$ and $yx_4 \in E$: 

\medskip

\noindent
Then $(x,y_1,x_2,y,x_4,y_3)$ induce a $2D-C_6$ which is impossible in this section.

\medskip

\noindent
{\bf Case 1.2} $xy_1 \in E$ and $yx_3 \in E$:

\medskip

\noindent
Then $(x,y_1,x_2,y,x_3,y_3)$ induce a $2D-C_6$ which is impossible in this section.

\medskip

\noindent
{\bf Case 1.3} $xy_2 \in E$ and $yx_4 \in E$:

\medskip

Then $(x,y_2,x_2,y,x_4,y_3)$ induce a $2D-C_6$ which is impossible in this section.

\medskip

Thus, assume that $xy_1 \notin E$ and $yx_4 \notin E$. Then $xy_2 \in E$ and $yx_3 \in E$. 
Since $(y_3,x,x_4,x_3,y,x_2,y_1,x_1)$ (with midpoint $y_3$) does not induce an $S_{1,1,5}$, we have $yx_1 \in E$. 
But then $(y_3,x,y_2,x_2,y,x_1)$ induce a $2D-P_6$ which is impossible in this section.
Thus, $y_1 \in D$ or $x_4 \in D$.

\medskip

\noindent
{\bf Case 2.} $y_1 \in D$ and $x_4 \notin D$:

\medskip

\noindent
Then $y_3$ must have a $D$-neighbor $x \in D$. 
Since $(y_3,x,x_4,x_3,y_2,x_2,y_1,x_1)$ (with midpoint $y_3$) does not induce an $S_{1,1,5}$, we have $xy_2 \in E$. 
But then $(x_1,y_1,x_2,y_2,x,y_3)$ induce a $2D-P_6$ which is impossible in this section. 

\medskip

\noindent
Analogously, $y_1 \notin D$ and $x_4 \in D$ is impossible. 
Thus, $y_1 \in D$ and $x_4 \in D$, and Lemma~\ref{P8secondinD} is shown.
\qed 

\medskip

In this section, by Lemma \ref{P8secondinD}, for every $P_8$ $P=(x_1,y_1,\ldots,x_4,y_4)$ in $G$, we have $y_1,x_4 \in D$. 
Then $D_{basis}$ can be updated by $D_{basis}:= D_{basis} \cup \{y_1,x_4\}$.
Moreover, since $y_2,x_3 \notin D$, $y_2$ and $x_3$ must have a $D$-neighbor, say $xy_2 \in E$ with $x \in D$ and $yx_3 \in E$ with $y \in D$. 

\medskip

Assume that $x_1,x_2,y_3,y_4 \in N_1$ and $y_2,x_3 \in N_2$. Then $x,y \in D \cap N_3$, and since there is no $2D-P_6$ in this section, $x$ does not have any neighbor in $N_3 \cup N_4$, and analogously, $y$ does not have any neighbor in $N_3 \cup N_4$. Thus, $x,y$ are $D_{basis}$-forced and 
$D_{basis}:= D_{basis} \cup \{y_1,x_4,x,y\}$, or it leads to a contradiction. 

\medskip

Thus, by every $P_8$ in $G$, $D_{basis}$ can be updated or it leads to a contradiction. 

\medskip

From now on, in the updated $D_{basis}$, there are only $P_8$'s in $N_0 \cup N_1$, and for every component $Q$ in $G \setminus (N_0 \cup N_1)$, $Q$ is $P_8$-free bipartite. Clearly, the e.d.s.\ problem can be done independently for every such component $Q$.  

\medskip

In \cite{BraMos2020P8}, we showed that ED is solvable in polynomial time for $P_8$-free bipartite graphs.    
Thus, Theorem \ref{EDS115frbipgr} is shown.

\medskip

\noindent
{\bf Acknowledgment.} The second author would like to witness that he just tries to pray a lot and is not able to do anything without that - ad laudem Domini.
 
\begin{footnotesize}

\end{footnotesize}

\end{document}